\documentclass[aps,prl,twocolumn,showpacs]{revtex4}
\usepackage{amssymb}
\usepackage{graphicx}
\usepackage{amsmath}
\usepackage{amsfonts}

\begin{document}

\title{Na$_{4}$Ir$_{3}$O$_{8}$ as a 3D spin liquid with fermionic spinons}
\date{\today}
\author{Yi Zhou$^{1,2}$, Patrick A. Lee$^3$, Tai-Kai Ng$^4$, and Fu-Chun Zhang$^1$ }
\affiliation{
Department of Physics, Center of Theoretical and Computational Physics, The University of Hong Kong, Hong Kong, China$^1$\\
Department of Physics and ITP, The Chinese University of Hong Kong, Hong Kong, China$^2$\\
Department of Physics, Massachusetts Institute of Technology, Cambridge, Massachusetts 02139, USA$^3$\\
Department of Physics, Hong Kong University of Science and Technology, Clear Water Bay Road, Kowloon, Hong Kong, China$^4$}
\begin{abstract}
Spin liquid states for spin-$1/2$ antiferromagnetic Heisenberg model on a hyperkagome lattice are studied. We
classify and study flux states according to symmetries. Applying this model to Na$_4$Ir$_3$O$_8$, 
we propose that the high temperature state may be described by a spinon Fermi surface, which forms a paired state with line nodes below 20~K.
The possible mixed spin singlet and spin triplet
pairing states are discussed according to the lattice symmetry which breaks inversion.
\end{abstract}
\maketitle

A spin liquid is a spin system where quantum fluctuations dominate its low energy behavior,
thereby a long range spin order can not be established even at zero temperature. 
A spin liquid is expected to have exotic properties such as spinons carrying $S=1/2$ excitations.  
After a long search, promising examples in two dimensions (2D) have been identified.\cite{Lee}
Recently a spinel related oxide, Na$_{4}$Ir$_{3}$O$_{8}$, was proposed as
the first candidate for a 3D spin liquid\cite{Takagi07}.
Temperature dependent spin susceptibility around room temperature yields an
antiferromagnetic (AFM) Curie-Weiss constant $\theta _{W}\sim 650$ K, and
there is no anomaly indicative of long range spin ordering down to $2$K.
Some kind of phase transition or cross-over seems to occur at a temperature $T_{c}\sim 20$%
K in that the specific heat $C_V$ divided by $T$ shows a rather sharp peak. On the other hand,  
the spin susceptibility $\chi\left( T\right) $ is
almost temperature independent. Using the experimental value of spin
susceptibility $\chi $ and specific heat ratio $\gamma $ at the specific
heat peak at $\sim 20K$, we find that the Wilson ratio $R_{W}=\frac{\pi^{2}k_{B}^{2}\chi}{3\mu _{B}^{2}\gamma}$ 
of the material is 0.88, which is
very close to that of a Fermi gas where $R_{W}$ is unity. Therefore, for a
wide range of temperature $T_{c}<T<\theta_W$, the system seems to be a
Fermi liquid of spinons. Below $T_{c}$ the specific heat decreases to zero as 
$C_V\sim T^{2}$, indicating a line nodal gap in the low lying quasiparticle
spectrum. However, this picture needs to be reconciled with the observation that 
the spin susceptibility $\chi $ remains almost constant.
In this letter, we will explain such phenomena based on a fermionic spin
liquid picture.

\begin{figure}[hpbt]
\includegraphics[width=8.6cm]{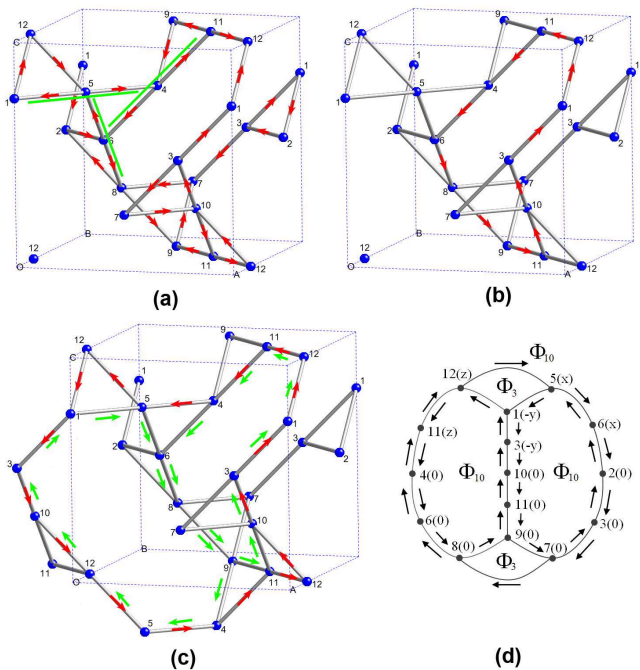}
\caption{(Color online) Hyperkagome lattices with space group $P4_132$.
 (a)~8 triangles giving rise to 3-site directed
loops. Straight green lines indicate a chiral pin-wheel. 
(b)~A 10-site directed loop.  Eleven other 10-site directed loops can be
generated from it by symmetry operations. (c)~A 16-site directed loop shown by red arrows. The green arrows show 
two oriented 10-site loops that are related by symmetry operations.  The figure shows 
how to decomposite the 16-site loop into sums of two 10-site
loops and four 3-site loops. (d)~An illustration of Eq.(\ref{phi3-10}),
where $\mu(i)$ denotes a site $\mu$ in unit cell $i$. We
unfold the lattice onto 2D and keep the topology, 
i.e. the surface enclosed by the third 10-site loop 
which makes up the rest of a closed surface is mapped to the rest of the 2D plane outside the figure.
Arrows indicate the orientation of three 10-site loops related by symmetry. 
The left 10-site loop is shown in 1(b).}
\label{hppkloop}
\end{figure}

The spins come from low spin $5d^{5}$ Ir$^{4+}$ ions and form a 3D network
of cornered shared triangles. This gives rise to a $S=1/2$ AFM coupled spin
system on a hyperkagome lattice (see Fig.\ref{hppkloop}). It should be
cautioned that because of the large atomic number, spin-orbit coupling in 
Ir is expected to be strong. 
This question was recently addressed by Chen and Balents.\cite{Balents}
We shall first discuss the case when spin-orbit coupling is negligible and comment on the strong coupling case later.

A hyperkagome lattice is a cubic lattice with 12 sites in a unit cell. There
are two types of hyperkagome lattices corresponding to space group $P4_{1}32$
and $P4_{3}32$ respectively. 
They differ just by chirality.\cite{Takagi07}
In this letter, we will focus on $P4_{1}32$. 
The chiral nature of the lattice is clearly seen as follows.  Each lattice site is connected to two bonds which forms a straight line 
(shown in green in Fig.1(a)).  Around each triangle, these straight lines define a chiral pinwheel.  
A set of arrows on each bond of the triangle can now be uniquely defined as shown in Fig.1(a).
The space group $P4_{1}32$ contains 24 symmetry operations; its corresponding point group is octahedral
group $O$\cite{crystallography}. 
It is readily seen that all the hyperkagome sites and bonds between nearest
neighbor sites are equivalent\cite{Balents}.
Since a unique oriented triangle is attached to each bond, all oriented triangles are also equivalent.  

We start with a nearest-neighbor (NN) Heisenberg model with AFM exchange $J$ and 
consider spin liquid states on such a hyperkagome lattice. The spin
liquid state is described by a fermionic trial wave function \cite{LeeNagaosaWen} which is the
ground state of a trial Hamiltonian, 
\begin{equation}
H_{trial}=H_{0}+H_{pair}  \label{Htrial}
\end{equation}
where $H_{0}$ is a tight binding Hamiltonian of spinons and $H_{pair}$
describes pairing between two spinons with 
\begin{eqnarray}
H_{0} &=&-\sum_{\left\langle i\mu ,j\nu \right\rangle,\alpha }t_{i\mu ,j\nu
}c_{i\mu\alpha }^{\dag }c_{j\nu\alpha }+h.c.,  \label{H0} \\
H_{pair} &=&-\sum_{\mathbf{k}\alpha \beta }\Delta _{\alpha \beta }(\mathbf{k}
)c_{\mathbf{k\alpha }}^{\dag }c_{-\mathbf{k\beta }}^{\dag }+h.c.,
\end{eqnarray}
where $i$ denotes unit cells, $\mu$ labels sites in a unit cell, 
$\left\langle i\mu ,j\nu \right\rangle $ denotes a pair of NN sites, 
$c_{i\mu }^{\dag }(c_{i\mu })$ denotes a fermionic spinon
creation (annihilation) operator, $\Delta _{\alpha \beta }(\mathbf{k})$ is a
gap function where $\alpha =\uparrow ,\downarrow $ are spin indices.
The spin liquid state is formed by Gutzwiller projection of the trial
wavefunction into a state with no double occupancy. $t_{i\mu ,j\nu }$ and
other parameters in the trial Hamiltonian should be viewed as variational
parameters obtained by minimizing the ground state energy through a
microscopic spin model. In this letter we treat the problem 
at the mean field level and consider spin liquid states
with full $P4_{1}32$ point group and translational symmetries.

First we consider $H_{0}$. The hopping integral $t_{i\mu ,j\nu }$ can be
written as $t_{i\mu ,j\nu }=\left\vert t\right\vert e^{-iA_{i\mu ,j\nu }}$
with $A_{i\mu ,j\nu }=-A_{j\nu,i\mu}$.
We shall identify all possible flux states with translational
and lattice symmetries.

All the loops on a hyperkagome lattice can be decomposed into two kinds of
\textquotedblleft elementary\textquotedblright\ loops, 
the 3-site and 10-site loops (see Fig.\ref{hppkloop}). 
As explained earlier, the 3-site loops are equivalent under symmetry operations
with the specified loop directions (see fig.(1)). 
Similarly with the 10-site loops, once the direction of one loop is fixed, symmetry operators will generate all 12 loops with specified directions.
Therefore all the 3-site and 10-site loops have same flux $\Phi _{3}$ 
and $\Phi_{10}$, respectively, for
states that respect translational and $P4_{1}32$ symmetry.

It is also helpful to define the 16-site loop
shown in Fig.1(c).  This loop forms a surface and repetition of this surface cuts through the whole lattice in the $xz$ plane.
All the 16-site loops form surfaces covering different
lattice planes and can be transformed to each other by symmetry operation.
The 16-site loops can be decomposed into sums of four 3-site loops and two 10-site
loops as shown in Fig.\ref{hppkloop}(c). 
Since the green arrows of one of the 10-site loops are opposed to the red arrows, the flux on the two 10-site loops cancel.  
Similarly, the $\Phi_3$'s also cancel, so that $\Phi_{16} = 0$.   We conclude that 
flux arrangements which satisfy the lattice symmetry do not permit a net flux through the lattice.  

\begin{figure}[htbp]
\includegraphics[width=8.4cm]{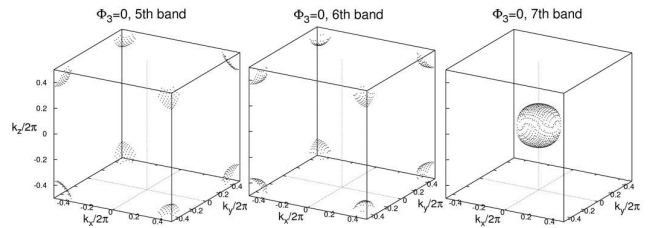} 
\caption{Fermi surfaces of the zero flux state at half filling.}
\label{FS1}
\end{figure}

Next we consider a volume enclosed by three 10-site loops and two 3-site loops as shown in Fig.1(d) 
projected to the 2D plane.  Since the total flux through the closed surface must be integer multiples of $2\pi$, we find
\begin{equation}
2\Phi _{3}=3\Phi _{10}( \text{mod}2\pi ).  \label{phi3-10}
\end{equation}
The states with zero and $\pi$ flux through the triangles are equivalent after projection.  
Moreover, the states with $\Phi_3$ and $\pi-\Phi_3$ are the same state after projection.  
This is because if we make a particle hole transformation $h_\uparrow = f_\downarrow^\dagger$, 
we have $t_{ij} \to -t_{ij}^\ast$, but the projection of the particle and hole states are 
the same at half-filling.\cite{Ran}  Thus we can restrict to $-\pi/2 \leq \Phi_3 < \pi/2$. There is 
only one time reversal invariant state $\Phi_3 = \Phi_{10} = 0$ which does not break lattice symmetry.
At half filling, Fermi surfaces of this state appear in 
only three bands as shown in Fig. \ref{FS1}. The other bands are either
fully filled or empty.

\begin{figure}[htbp]
\includegraphics[width=6.4cm]{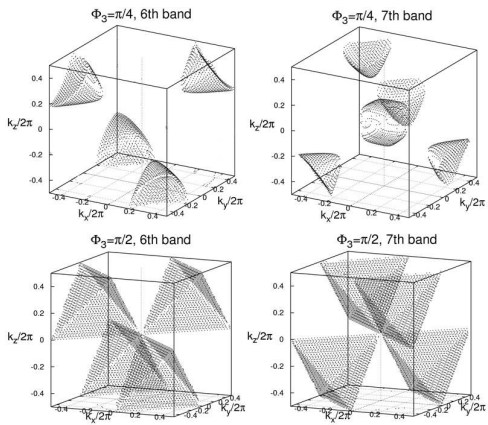} 
\caption{Fermi surfaces of flux states with broken time reversal symmetry at
half filling.}
\label{FS2}
\end{figure}

Breaking time reversal symmetry will lead to flux states with arbitrary fluxes 
$\Phi _{3}$ and $\Phi_{10} $ satisfying Eq.(\ref{phi3-10}), and still requires 
$\Phi _{16}=0$. All of these states, with or without time reversal symmetry,
have finite Fermi surfaces at half filling (see Figs. \ref{FS1} and \ref{FS2}).

To determine what is the most stable state, we perform a mean
field theory calculation on the AFM Heisenberg model, $H=J\sum_{\langle i\mu
,j\nu \rangle }\vec{S}_{i\mu }\cdot\vec{S}_{j\nu }$ with trial Hamiltonian $%
H_{trial}=H_0$, first neglecting the pairing terms. By standard $U\left(
1\right) $ slave boson mean field theory, we have the relation 
$t= {3\over8} J\left\vert \sum_{\sigma }\langle c_{i\mu \sigma }^{\dagger} c_{j\nu
\sigma }\rangle \right\vert$ and the ground state energy is 
$E_{0}/N_{b}=-8t^{2}/3J$, where $N_{b}$ is the number of bonds. We find that 
$E_{0}$ reaches its minimum at $\Phi_{3}=0$.

Phenomenologically we expect that $H_0$ determines the physics of the spin
liquid state at $T>T_c$ and a spinon pairing gap characterized by $H_{pair}$
is opened up at $T<T_c$. The power law behavior $C_{V}\propto T^{2}$
observed at $T<T_{c}$ indicates that the gap has line nodes
on Fermi surfaces. For triplet pairing, 
group theoretical analysis tells us that a spin triplet pairing state on a cubic lattice can
create only full or point nodal gaps\cite{Sigrist91}. 
This would seem to imply singlet pairing. However, because of broken inversion symmetry on 
a hyperkagome lattice\cite{crystallography}, spin singlet and triplet pairing states can now
mix together in the presence of spin orbit coupling\cite{Inversion}. In terms of $d$-vector the gap function 
$\Delta _{\alpha \beta }(\mathbf{k})$ can be written in matrix form\cite{Leggett}, 
\begin{equation}
\Delta (\mathbf{k})=i\left( d_{0}\left( \mathbf{k}\right) \sigma _{0}+%
\mathbf{d}\left( \mathbf{k}\right) \cdot \mathbf{\sigma }\right) \sigma _{y},
\label{delta}
\end{equation}%
with the energy dispersion of quasiparticles 
\begin{equation}
E_{\mathbf{k}}=\sqrt{\xi _{\mathbf{k}}^{2}+d_{0}d_{0}^{\ast }+\mathbf{d}
\cdot \mathbf{d}^{\ast }\pm \sqrt{|\mathbf{d}\times \mathbf{d}^{\ast
}|^{2}+\left( d_{0}^{\ast }\mathbf{d}+d_{0}\mathbf{d}^{\ast }\right) ^{2}}},
\label{Emix}
\end{equation}
where $\xi _{\mathbf{k}}=\varepsilon _{\mathbf{k}}-\mu $, $\varepsilon _{%
\mathbf{k}}$ is the energy dispersion of spinons calculated from $H_{0}$ and 
$\mu $ is the chemical potential. We have assumed $\xi _{-\mathbf{k}}=\xi _{\mathbf{k}}$ 
in deriving $E_{\mathbf{k}}$,
which is valid at half filling in
presence of time-reversal symmetry. The unitary condition\cite{Leggett} is
given by $i\mathbf{d}\times \mathbf{d}^{\ast }+d_{0}^{\ast }\mathbf{d}+d_{0}%
\mathbf{d}^{\ast }=0$. It is obvious that the mixing of singlet and triplet
states leads to nonunitary states even if the $d$-vector is real. For a
nonunitary state, there are two branches of $E_{\mathbf{k}}$ with the energy
gap determined by the lower branch. We will see how such mixing may lead to
nodal line states later. 
In such pairing states, the $U(1)$ gauge symmetry 
is broken by the spinon pairing $\langle c_{i\sigma}^\dagger c_{j\sigma^\prime}^\dagger \rangle$. 
However, a local $Z_2$ symmetry, where $c_{i\sigma} \rightarrow \pm c_{i\sigma}$ together 
with a sign change in the gauge variable on the link remains intact, 
so that they can be classified as $Z_2$ spin liquids.\cite{Wen}

Spin pairing states on a cubic lattice with inversion symmetry are
classified in Ref.\cite{Sigrist91}. What we need to do is to mix the pairing
states as a linear combination of odd parity (spin singlet) and even parity (spin triplet)
pairing states within an irreducible representation $\Gamma $ of
octahedral group $O$. For simplicity, we shall only consider $s$-wave,
(or extended $s$-wave), $p$-wave and $d$-wave states.
In this case the available possibilities are $\Gamma _{1}$ ($s+p$-wave), $%
\Gamma _{3}$ ($d+p$-wave) and $\Gamma _{5}$ ($d+p$-wave). We shall also
limit ourselves to spin rotation invariant states with $\left\langle\mathbf{S%
}\right\rangle =0$ for simplicity. Therefore $\mathbf{d}\left( \mathbf{k}%
\right)\times \mathbf{d}^{\ast }\left( \mathbf{k}\right) =0$ and $\mathbf{d}%
\left( \mathbf{k}\right) $ is real apart from a overall phase factor. The
lower energy gap is then $\Delta_{\vec{k}}=\sqrt{d_{0}d_{0}^{\ast }+\mathbf{d%
}\cdot \mathbf{d}^{\ast }-\left\vert d_{0}^{\ast }\mathbf{d} +d_{0}\mathbf{d}%
^{\ast }\right\vert }$ and results in a nodal surface $(\Delta_{\vec{k}}=0)$
in $\vec{k}$-space because of mixing between different parity states as we
shall see in the following. A nodal line of gap in quasi-particle excitation
spectrum is produced by the cross line between nodal surface 
and Fermi surface ($\xi_{\vec{k}}=0$).

First we consider 1D representation $\Gamma _{1}$. In this case the
spin singlet part $d_{0}$ may be $s$-wave or extended $s$-wave, with a $p$%
-wave state for spin triplet part $\mathbf{d}$. For $s$-wave, we have $d_{0}(%
\mathbf{k})=\Delta _{s}$ and $\mathbf{d}(\mathbf{k)}=\Delta _{t}(\sin k_{x}%
\hat{x}+\sin k_{y}\hat{y}+\sin k_{z}\hat{z})$. $\Delta _{\vec{k}}=0$ implies 
$\sin ^{2}k_{x}+\sin ^{2}k_{y}+\sin ^{2}k_{z}=|\frac{\Delta _{s}}{\Delta _{t}}|^{2}$
which produces a nodal surface. Notice that the nodal surface shrinks to a
point when $\Delta _{s}\to 0$. For extended $s$-wave, we have $%
d_{0}(\mathbf{k})=\Delta _{s}(\cos k_{x}+\cos k_{y}+\cos k_{z})$ and $%
\mathbf{d}(\mathbf{k)}=\Delta _{t}(\sin k_{x}\hat{x}+\sin k_{y}\hat{y}+\sin
k_{z}\hat{z})$. The node surface is given by 
$|\frac{\Delta_{s}}{\Delta_{t}}|^{2}=\frac
{\sin ^{2}k_{x}+\sin ^{2}k_{y}+\sin ^{2}k_{z}}{(\cos k_{x}+\cos k_{y}+\cos k_{z})^{2}}$.

Similar analysis can be applied to 2D representation $\Gamma _{3}$ and
3D representation $\Gamma _{5}$. For representation $\Gamma _{3}$ ($d+p$-wave), we have
$d_{0}(\mathbf{k})=\Delta _{s}[2u_{1}\cos
k_{z}-(u_{1}+\sqrt{3}u_{2})\cos k_{x}-(u_{1}-\sqrt{3}u_{2})\cos k_{y}]$ and 
$\mathbf{d}(\mathbf{k)}=\Delta _{t}[2v_{1}\sin k_{z}\hat{x}-(v_{1}+\sqrt{3}%
v_{2})\sin k_{x}\hat{y}-(v_{1}-\sqrt{3}v_{2})\sin k_{y}\hat{z}]$, where $%
u_{1}^{2}+u_{2}^{2}=v_{1}^{2}+v_{2}^{2}=1$ with $u_{1(2)}$ and $v_{1(2)}$
being real. For representation $\Gamma _{5}$ ($d+p$-wave), we have
$d_{0}(\mathbf{k})=\Delta _{s}(u_{x}\sin k_{y}\sin k_{z}+u_{y}\sin
k_{z}\sin k_{x}+u_{z}\sin k_{x}\sin k_{y})$ and $\mathbf{d}(\mathbf{k)}%
=\Delta _{t}[v_{x}(\sin k_{z}\hat{y}+\sin k_{y}\hat{z})+v_{y}(\sin k_{y}\hat{%
z}+\sin k_{z}\hat{x})+v_{z}(\sin k_{y}\hat{x}+\sin k_{x}\hat{y})]$, where $%
u_{x}^{2}+u_{y}^{2}+u_{z}^{2}=v_{x}^{2}+v_{y}^{2}+v_{z}^{2}=1$, with $%
u_{x(y,z)}$ and $v_{x(y,z)}$ being real. For all these states it is possible to have
nodal surfaces which will shrink to points as $\Delta _{s}\to 0$.
Thus we conclude that the pairing must be singlet or singlet with triplet admixture 
due to spin-orbit coupling in order to have node lines. 

Next we consider the spin susceptibility of such mixed states. We
neglect spin-orbit interaction and consider $\vec{S}$ as `real' spins at first.
Singlet coupling predicts the spin susceptibility will go towards zero below 
$T_c$, which is not observed experimentally.  If both singlet and triplet are present 
and spin-orbit scattering is weak (in the sense to be defined later), we find the $k$-dependent electronic
contribution to spin susceptibility is given by $\frac{\chi _{ii}(\mathbf{k})}{\chi
_{N}(\mathbf{k})}= 1-\frac{d_{0}d_{0}^{\ast }+d_{i}^{\ast }d_{i}}{%
d_{0}d_{0}^{\ast }+\mathbf{d\cdot d}^{\ast }}+ \frac{d_{0}d_{0}^{\ast
}+d_{i}^{\ast }d_{i}}{d_{0}d_{0}^{\ast }+\mathbf{d\cdot d}^{\ast }}Y(\mathbf{%
k};T)$, where $i=x,y,z$, and $Y(\mathbf{k};T)$ is the $k$-dependent Yosida
function\cite{Leggett}. 
Assuming that the $\vec{d}$ vector is pinned to the lattice, 
for a polycrystalline sample we should average over
all spatial directions, resulting in 
${\frac{\chi _{s}}{\chi _{N}}}=\frac{2}{3}-\frac{2}{3}\frac{\left\vert
d_{0}\right\vert ^{2}} {\left\vert d_{0}\right\vert^{2}+\left\vert \mathbf{d}
\right\vert ^{2}}+(\frac{1}{3}+\frac{2}{3}\frac{ \left\vert d_{0}\right\vert
^{2}}{\left\vert d_{0}\right\vert^{2}+\left\vert \mathbf{d}\right\vert ^{2}})Y(T)$,
where $Y\left( T\right)$ is the Yosida function, $\chi _{s}$ is the spin
susceptibility below $T_{c}$ and $\chi _{N}$ is the Pauli spin
susceptibility at the normal state. Therefore $\frac{\chi _{s}}{\chi _{N}}$ reduces
to $\frac{2}{3}-\frac{2}{3}\frac{\left\vert d_{0}\right\vert ^{2}}{%
\left\vert d_{0}\right\vert^{2}+ \left\vert \mathbf{d}\right\vert ^{2}}$ at
zero temperature. If the spin triplet pairing dominates, $\frac{\chi _{s}}{\chi _{N}}
\to \frac{2}{3}$; if the spin singlet pairing dominates, $\frac{\chi _{s}}{\chi _{N}}\to 0$. 
Neither of these cases are observed in experiment, where $\chi$ changes little below $T_c$.

In order to explain the absence of change in $\chi$ below 20~K, we appeal to the effect of spin-orbit coupling. 
It is well known that in conventional BCS singlet superconductors, the Knight shift hardly changes below $T_c$ 
for heavy elements such as Sn and Hg.\cite{Adroes}  It was quickly realized that this is due to 
the destruction of spin conservation due to the spin-orbit coupling.  A clear explanation was given by 
Anderson \cite{Anderson}, who introduced the notion of time reversed pairing states.  Let us consider the 
imaginary part of the spin response function $\chi^{\prime\prime} (q, \omega)$.  If total spin is conserved, the 
dynamics is diffusive and $\chi^{\prime\prime} (q, \omega)$ will have a central peak in $\omega$ space with 
width $Dq^2$ which goes to zero as $q \to 0$.  Superconductivity gaps out all low frequency excitations, 
thus removing this central peak.  By Kramers-Krong relation the real part $\chi^\prime (q=0, \omega=0)$ vanishes 
in the superconducting ground state.  In the presence of spin-orbit coupling, the total spin is not conserved, 
but decays with a lifetime $\tau_s$.  In this case $\chi^{\prime\prime}(q=0,\omega)$ has a central peak with 
a width ${1\over \tau_s}$. The superconducting gap $\Delta$ cuts a hole in the $\chi^{\prime\prime}(\omega)$ 
for $\omega < \Delta$, but leaves the region $\omega \gg \Delta$ intact, in agreement with physical expectation 
that the high frequency region should be unaffected by pairing.  
By Kramer-Kronig relation, $\chi^\prime$ will be reduced, but if the spin-orbit coupling is sufficiently strong such that
\begin{equation}
{1\over \tau_s} \gg \Delta \label{delta-tau}
\end{equation}
the reduction will be small, i.e.
$
{\chi_s \over \chi_N} = 1-{\cal O} (\Delta \tau_s) .
$
Equation (\ref{delta-tau}) is the criterion needed to have very little charge in the spin susceptibility below $T_c$.

Let us now discuss to what extent our model may be applicable to Na$_4$Ir$_3$O$_8$.  The $t_{2g}$ levels are split 
by crystal fields due to the presence of Na by an amount of order $E_3$. Chen and Balents \cite{Balents} made the 
important observation that the $t_{2g}$ levels may be considered as a $\vec{L} = 1$ multiplet and distinguish between 
the strong and weak coupling limits, depending on the spin-orbit energy $\lambda$.  In the strong coupling case
\begin{equation}
\lambda \gg E_3, \label{l-e3}
\end{equation}
the spin and orbital degrees of freedom lock to form $\vec{J} = \vec{L}+\vec{S}$ and it turns out that 
$J = {1\over 2}$. In this case, the direct exchange between the Ir ions gives an \emph{isotropic} Heisenberg 
model $J \vec{S}_i \cdot \vec{S}_j$, if $\vec{S}$ is interpreted as $\vec{J}$. The $g$ factor is 2 while 
superexchange via oxygen gives rise to anisotropy and Dzyaloshinski-Moriya (DM) terms.  
Since the experimentally measured $g$ coupling is close to 2, 
Chen and Balents concluded that 
the Na$_4$Ir$_3$O$_8$ system must be either close to weak coupling or strong coupling 
in the sense of Eq.(\ref{l-e3}), and strong coupling seems more likely, given the size of $\lambda$.  
Provided that the exchange path is dominated by direct exchange, our starting point with 
the isotropic Heisenberg model and the fermion representation model remains valid, provided that 
the spin labels in Eq.(2) are to be understood as $J_z = \pm {1 \over 2}$.  The correction terms 
are anisotropy and DM terms which are expected to be of order $(E_3/\lambda)^2 \approx (g-2)^2$ in this case. 
These correction terms spoil the conservation of total $J$, giving rise to a decay rate analogous to ${1 \over \tau_s}$. 
In the limit $\Delta\tau_s \ll 1$ we again expect that the spin susceptibility is not much affected by the onset of pairing. 
This may be the case closest to experiment.  We emphasize that the criterion for strong or weak spin-orbit coupling given 
by Eq.(\ref{l-e3}) is different from Eq.(\ref{delta-tau}) which determines the magnitude of the drop in $\chi$.

In summary, we studied fermionic spin liquid states described
by trial Hamiltonian\ (\ref{Htrial}) on a 3D hyperkagome lattice,
and classified all possible flux states, which all have Fermi surfaces.
The broken inversion symmetry may lead to the mixing of spin
singlet and triplet states in the spinon paired states, resulting in formation of line nodal gaps,
which provides a natural explanation to the observed low temperature
specific heat $C_{V}\sim T^{2}$ in Na$_4$Ir$_3$O$_8$. The strong spin-orbit coupling
explains to the observed spin susceptibility, which is more or less temperature independent at
temperatures below $\sim T_{c}$.  The transition at 20~K is then interpreted as a transition between 
a $U(1)$ spin liquid to a $Z_2$ spin liquid where spinons are paired at low temperature.
This theory also predicts the appearance of superconductivity if the Na$_4$Ir$_3$O$_8$ system can be doped.

We thank L. Balents, Y.B. Kim, H. Takagi and especially T. Senthil for helpful discussions. 
YZ, TKN and FCZ acknowledge HKSAR RGC grants for partial financial support. 
PAL acknowledges the support of DOE grant DE--FG02--03ER46076 and thanks 
the Institute for Advanced Studies at HKUST for its hospitality.

\end{document}